\pdfoutput=1
\documentclass[a4paper,11pt]{article}
\usepackage{pos}
\usepackage{colortbl,tabularx}
\usepackage{placeins}
\newcommand{\red}[1]{{\textcolor{red}{#1}}}

\newcommand{\wm}{\phantom{-}}

\title{Status of next-generation $\Lambda_b \to p, \Lambda, \Lambda_c$ form-factor calculations}

\author*{Stefan Meinel}

\affiliation{Department of Physics, University of Arizona, Tucson, AZ 85721, USA}

\emailAdd{smeinel@arizona.edu}

\abstract{I present preliminary results of next-generation lattice-QCD calculations of the $\Lambda_b \to p$, $\Lambda_b \to \Lambda$, and $\Lambda_b \to \Lambda_c$ form factors based on RBC/UKQCD gauge-field ensembles with 2+1 flavors of domain-wall fermions. Compared to the work published in 2015 and 2016, the new calculations include three additional ensembles (one with 139 MeV pion mass, one with 0.073 fm lattice spacing, and one with another volume) and were performed with a more accurate tuning of the charm and bottom anisotropic clover action parameters.}

\FullConference{The 40th International Symposium on Lattice Field Theory (Lattice 2023)\\
July 31st - August 4th, 2023\\
Fermi National Accelerator Laboratory\\}

\begin{document}
\maketitle

\section{Introduction}

Semileptonic decays of heavy baryons provide determinations of CKM matrix elements, constraints on flavor-changing neutral-current Wilson coefficients, and tests of lepton-flavor universality. The nonzero spins of the initial and final-state baryons make the decay amplitudes sensitive to all possible operator structures appearing in the weak effective Hamiltonian and allow a large number of angular observables that can be used to disentangle these structures. The most important charged-current bottom-baryon semileptonic decays are $\Lambda_b \to p \ell^- \bar{\nu}$ and $\Lambda_b \to \Lambda_c \ell^- \bar{\nu}$, which have been analyzed by LHCb to determine $|V_{ub}/V_{cb}|$ \cite{LHCb:2015eia} and test $\tau$-versus-$\mu$ lepton flavor universality \cite{LHCb:2022piu}. The most important bottom-baryon FCNC decays are $\Lambda_b \to \Lambda \ell^+\ell^-$ and $\Lambda_b \to \Lambda \gamma$. LHCb measurements of the $\Lambda_b \to \Lambda \mu^+\mu^-$ differential branching fraction and $\Lambda_b \to \Lambda(\to p\pi^-) \mu^+\mu^-$ angular observables \cite{LHCb:2015tgy,LHCb:2018jna} have been used to constrain the $b\to s\mu^+\mu^-$ Wilson coefficients $C_{9,10}^{(\prime)}$ \cite{Blake:2019guk}.

The theoretical description of these decay processes involves hadronic matrix elements of the form $\langle F | \bar{q}\Gamma b |\Lambda_b\rangle $, where $F$ is the baryon in the final state. These matrix elements can be expressed in terms of form factors $f_0$, $f_+$, $f_\perp$ (for $\Gamma=\gamma^\mu$), $g_0$, $g_+$, $g_\perp$ (for $\Gamma=\gamma^\mu\gamma_5$), and $h_+$, $h_\perp$, $\widetilde{h}_+$, $\widetilde{h}_\perp$ (for $\Gamma=\sigma_{\mu\nu}$) \cite{Feldmann:2011xf}, which are functions of the four-momentum-transfer-squared, $q^2$. The phenomenological applications cited above used lattice-QCD calculations of these form factors that were published in 2015 and 2016 \cite{Detmold:2015aaa,Detmold:2016pkz}. For the $|V_{ub}/V_{cb}|$ determination, the 2015 experimental and lattice uncertainties were about the same size, approximately 5\% each \cite{LHCb:2015eia,Detmold:2015aaa}. The LHCb measurement is projected to reach 2\% precision with 50 ${\rm fb}^{-1}$ of integrated luminosity \cite{Albrecht:2017odf}, requiring a commensurate improvement of the lattice-QCD calculation. For the $\Lambda_b \to \Lambda \mu^+\mu^-$ differential branching fraction at low $q^2$, the 2015 experimental uncertainty \cite{LHCb:2015tgy} was already smaller than the uncertainty of the 2016 Standard-Model prediction using the form factors from lattice QCD \cite{Detmold:2016pkz}, and in the meantime LHCb has accumulated much more data \cite{DiCanto:2022icc}. In summary, higher-precision lattice-QCD determinations of the $\Lambda_b \to p$, $\Lambda_b \to \Lambda$, and $\Lambda_b \to \Lambda_c$ form factors are needed. I will report here on work underway toward this goal.

\section{Parameters of the next-generation calculations}

The ongoing next-generation calculations of the $\Lambda_b \to p, \Lambda, \Lambda_c$ form factors use the same types of lattice actions as the 2015/2016 calculations: the Iwasaki action for the gluons, domain-wall actions for the $u$, $d$, and $s$ quarks, and anisotropic clover actions for the $b$ and $c$ quarks. The new calculations include three additional ensembles gauge-field configurations generated by the RBC and UKQCD collaborations \cite{RBC:2014ntl,Boyle:2018knm} and incorporate several other improvements as discussed in the following.

\begin{table}
\begin{center}
\footnotesize
\begin{tabular}{|llcllllcccc|}
\hline
   \rowcolor{gray!10}  Label & $N_s^3\times N_t$ & $\beta$ & $a m_{u,d}^{(\rm sea)}$ \hspace{-2ex} & $a m_{u,d}^{(\rm val)}$ \hspace{-2ex} & $a m_s^{(\rm sea)}$ \hspace{-2ex} & $a m_s^{(\rm val)}$ \hspace{-2ex} &  $a\:[\rm fm]$ & \hspace{-2ex} $m_\pi^{(\rm sea)}$ [MeV]  & \hspace{-2ex} $m_\pi^{(\rm val)}$ [MeV] \hspace{-2ex}  & Samples  \\
\hline
C14 & $24^3\times 64$   & 2.13   & 0.005     & \red{0.001} & 0.04  & 0.04     & $\approx0.111$       & $\approx340$  & \red{$\approx250$}  & 2672    \\
C24 & $24^3\times 64$   & 2.13   & 0.005     & \red{0.002} & 0.04 & 0.04      & $\approx0.111$       & $\approx340$  & \red{$\approx270$}  & 2676    \\
C54 & $24^3\times 64$   & 2.13   & 0.005     & 0.005 & 0.04 & 0.04      & $\approx0.111$       & $\approx340$  & $\approx340$   & 2782   \\
C53 & $24^3\times 64$   & 2.13   & 0.005     & 0.005 & 0.04 & \textcolor{purple}{0.03}      & $\approx0.111$       & $\approx340$  & $\approx340$   & 1205   \\
F23 & $32^3\times 64$   & 2.25   & 0.004     & \red{0.002} & 0.03  & 0.03      & $\approx0.083$       & $\approx304$  & \red{$\approx230$} & 1907     \\
F43 & $32^3\times 64$   & 2.25   & 0.004     & 0.004 & 0.03  & 0.03      & $\approx0.083$       & $\approx304$  & $\approx304$     & 1917  \\
F63 & $32^3\times 64$   & 2.25   & 0.006     & 0.006 & 0.03  & 0.03      & $\approx0.083$       & $\approx361$  & $\approx361$     & 2782   \\
\hline
\end{tabular}

\vspace{1ex}

\begin{tabularx}{\linewidth}{|XXcXXXccc|}
\hline
   \rowcolor{gray!10}  Label & $N_s^3\times N_t$ & $\beta$ & $a m_{u,d}^{(\rm sea,val)}$  & $a m_s^{(\rm sea)}$  & $a m_s^{(\rm val)}$ & $a\:[\rm fm]$ &  $m_\pi^{(\rm sea,val)}$ [MeV] & Samples  \\
\hline
CP     & $48^3\times 96$   & 2.13   & 0.00078   & 0.0362  & 0.0362                       & $\approx0.114$  & $\approx139$   & 80 ex,  2560 sl \\
C005LV & $32^3\times 64$   & 2.13   & 0.005     & 0.04    & \textcolor{purple}{0.0323}   & $\approx0.111$  & $\approx340$   & 186 ex,  5022 sl \\
C005   & $24^3\times 64$   & 2.13   & 0.005     & 0.04    & \textcolor{purple}{0.0323}   & $\approx0.111$  & $\approx340$   & 311 ex,  4976 sl \\
F004   & $32^3\times 64$   & 2.25   & 0.004     & 0.03    & \textcolor{purple}{0.0248}   & $\approx0.083$  & $\approx304$   & 251 ex,  4016 sl \\
F006   & $32^3\times 64$   & 2.25   & 0.006     & 0.03    & \textcolor{purple}{0.0248}   & $\approx0.083$  & $\approx361$   & 223 ex,  3568 sl \\
F1M    & $48^3\times 96$   & 2.31   & 0.002144  & 0.02144 &  \textcolor{purple}{0.02217} & $\approx0.073$  & $\approx232$   & 113 ex,  3616 sl \\
\hline
\end{tabularx}
\end{center}
\caption{\label{tab:ensembles}The data sets used for the 2015/2016 calculations \cite{Detmold:2015aaa,Detmold:2016pkz} (upper table) and for the ongoing next-generation calculations (lower table).  }
\end{table}

For reference, the upper part of Table \ref{tab:ensembles} shows the data sets used in the 2015/2016 calculations. These data sets in fact come from three independent ensembles only. Partial quenching with $m_{u,d}^{(\rm val)}<m_{u,d}^{(\rm sea)}$ (highlighted in red color in the table) was used to generate the C14, C24, and F23 data sets to reach lower valence pion masses. This choice, however, resulted in small values of $m_\pi^{(\rm val)}L$ that made finite-volume errors the dominant systematic uncertainty in the ratio of decay rates relevant for the $|V_{ub}/V_{cb}|$ determination \cite{Detmold:2015aaa}. The data sets used for the next-generation calculations are shown in the lower part of Table \ref{tab:ensembles}. All six data sets now come from different ensembles and all use $m_{u,d}^{(\rm val)}=m_{u,d}^{(\rm sea)}$, satisfying $m_\pi L\gtrsim 4$. The CP ensemble has an approximately physical pion mass and the F1M ensemble has a third, finer lattice spacing; these two ensembles use a M\"obius domain-wall action instead of the Shamir action used otherwise \cite{RBC:2014ntl,Boyle:2018knm}. The C005LV ensemble differs from C005 only in the lattice size. The valence strange-quark masses (highlighted in purple in the table where different from $m_{s}^{(\rm sea)}$) are now tuned to the physical values. I newly computed all quark propagators using all-mode averaging \cite{Shintani:2014vja} with low-mode deflation for the light quarks; the numbers of exact and sloppy samples are annotated with ``ex'' an ``sl'' in the table. The Gaussian source smearing now uses APE-smeared gauge links, and the smearing width for the F004 and F006 data sets was increased compared to the 2015/2016 calculations to now match the width used on the C005/C005LV ensembles in physical units.

The actions used for the bottom and charm quarks have the form
\begin{equation}
 S_Q = a^4 \!\sum_x \bar{Q} \bigg[ {m_Q} + \gamma_0 \nabla_0 - \frac{a}{2} \nabla^{(2)}_0 + {\nu}\!\sum_{i=1}^3\left(\gamma_i \nabla_i - \frac{a}{2} \nabla^{(2)}_i\right)
- {c_E}  \frac{a}{2}\! \sum_{i=1}^3 \sigma_{0i}F_{0i} - {c_B} \frac{a}{4}\! \sum_{i,\, j=1}^3 \! \sigma_{ij}F_{ij}  \bigg] Q.
\end{equation}
By appropriately tuning the parameters $am_Q$, $\nu$, $c_E$, and $c_B$ for each lattice spacing, heavy-quark discretization errors that scale as powers of $am_Q$ can be removed \cite{El-Khadra:1996wdx,Chen:2000ej,Aoki:2001ra,Aoki:2003dg,Christ:2006us,Lin:2006ur}. In the 2015/2016 $\Lambda_b$ calculations, I used the bottom-quark parameters from Ref.~\cite{RBC:2012pds}\footnote{Reference \cite{RBC:2012pds} uses the notation $m_0=m_Q$, $\zeta=\nu$ and $c_P=c_{E}=c_{B}$.}, which were tuned nonperturbatively using the $B_s^{(*)}$ dispersion relation and hyperfine splitting, and charm-quark parameters from Ref.~\cite{Brown:2014ena}, where the mass and anisotropy were tuned using the charmonium dispersion relation and $c_{E}$, $c_B$ were set to the tadpole-improved tree-level values. For the second-generation $\Lambda_b$ calculations, I performed a new tuning of the heavy-quark parameters, using updated lattice-spacing determinations from RBC/UKQCD \cite{RBC:2014ntl,Boyle:2018knm} and using the heavy-strange dispersion relation and hyperfine splitting for both bottom and charm. The $B_s$ or $D_s$ dispersion relation is written as $\displaystyle (aE(\mathbf{p}))^2 \approx (aM_{\rm rest})^2 + c^2 (a\mathbf{p})^2,$ where $c^2=M_{\rm rest}/{M_{\rm kin}}$. I calculated $c^2$ directly as $c^2 = [(aE(\mathbf{p}))^2 - (aE(\mathbf{0}))^2]/(a\mathbf{p})^2$ with $\mathbf{p}^2=1\cdot(2\pi/L)^2$ and confirmed that higher momenta give consistent results. The tuning conditions are the same as in Ref.~\cite{RBC:2012pds}: $E_{H_s}(\mathbf{0})=M_{H_s,{\rm exp}}$, $c^2_{H_s}=1$, and $E_{H_s^*}(\mathbf{0})-E_{H_s}(\mathbf{0})=M_{H_s^*,{\rm exp}}-M_{H_s,{\rm exp}}$, where the experimental values are taken from Ref.~\cite{ParticleDataGroup:2022pth}. My results for these quantities for several choices of the heavy-quark parameters are shown in Table \ref{tab:RHQparams}. The final choice for $am_Q$, $\nu$, $c_{E}=c_{B}$ (highlighted in green) on each ensemble was obtained using a linear fit to the other results and solving the tuning conditions for the parameters. Runs with these final parameters then confirmed that the tuning conditions are indeed satisfied.

\begin{table}

\vspace{-3ex}

\begin{center}
\scriptsize
\begin{tabular}{|lllllllll|}
  \hline
   \rowcolor{gray!15}  Ens. \hspace{1.8ex}  & $am_Q$ \hspace{1.8ex} & $\nu$ & $c_{E,B}$  & $a E_{B_s}$  \hspace{1.5ex}     & $E_{B_s}$ [GeV]    & $c^2_{B_s}$        &  $a E_{B^*_s}-a E_{B_s}$ \hspace{0.3ex} &  $E_{B^*_s}-E_{B_s}$ [MeV]\!\!\!\null  \\
  \hline
   CP  &  8.45     &  3.1    &  5.8   & 3.1084(13)  &  5.376(12)  & 0.914(13)   & 0.0288(13)  & 49.8(2.3)   \\ 
  \rowcolor{green!20}  CP & $8.1476$ & $3.3743$ & $5.3944$  & 3.1025(12)  &  5.366(12)  & 1.001(14)   & 0.0282(13)  & 48.8(2.3)   \\
  \hline
   C005   & 7.42     &  2.92   &  4.86 & 3.00324(81) & 5.360(15)   & 0.9184(89) & 0.02602(52) & 46.44(94)   \\ 
  \rowcolor{blue!10} C005   & 7.471    &  2.929  &  4.92 & 3.00801(81) & 5.369(15)   & 0.9184(89)   & 0.02616(53) & 46.69(95)   \\ 
   C005   & 7.9      &  2.929  &  4.92 & 3.05463(85) & 5.452(15)   & 0.8895(91)  & 0.02482(53) & 44.29(96)   \\ 
   C005   & 7.471    &  2.929  &  6    & 2.97872(77) & 5.316(15)   & 0.9399(85)  & 0.03165(58) & 56.5(1.1)   \\ 
   C005   & 7.471    &  3.3    &  4.92  & 3.03049(75) & 5.409(15)   & 1.0186(89)  & 0.02665(48) & 47.56(86)   \\ 
   \rowcolor{green!20}   C005   & $7.3258$ & $3.1918$ & $4.9625$ & 3.00695(75) & 5.367(15)   & 1.0013(88)  & 0.02719(49) & 48.52(88)   \\ 
  \hline
  \rowcolor{blue!10} F004   & 3.485    & 1.76   &  3.06  & 2.25010(82) & 5.363(19)   & 0.855(12)  & 0.02053(62) & 48.9(1.5)   \\
   F004   & 3.3      & 1.76   &  3.06  & 2.20958(79) & 5.266(19)   & 0.877(12)   & 0.02148(61) & 51.2(1.5)   \\
   F004   & 3.485    & 1.76   &  3.7   & 2.21551(78) & 5.280(19)   & 0.877(12)  & 0.02468(68) & 58.8(1.6)   \\
   F004   & 3.485    & 2.2    &  3.06  & 2.29690(70) & 5.474(20)   & 1.047(12)   & 0.02126(51) & 50.7(1.2)   \\
    \rowcolor{green!20}  F004   & $3.2823$ & $2.0600$ & $2.7960$ & 2.25175(71) & 5.366(19)   & 1.004(12)    & 0.02046(51) & 48.8(1.2)   \\
  \hline
   F1M    & 2.4538   & 1.7563   & 2.6522   & 1.97455(85) & 5.347(20) & 0.960(16) & 0.01876(79) & 50.8(2.2) \\ 
   F1M    & 2.55     & 1.7563   & 2.6522   & 2.00012(88) & 5.416(20) & 0.943(17) & 0.01821(81) & 49.3(2.2) \\ 
   F1M    & 2.4538   & 2        & 2.6522   & 2.00698(75) & 5.435(20) & 1.089(16) & 0.01926(70) & 52.2(1.9) \\ 
   F1M    & 2.4538   & 1.7563   & 3.1      & 1.94472(81) & 5.266(20) & 0.983(16) & 0.02151(83) & 58.3(2.3) \\ 
    \rowcolor{green!20}  F1M    & $2.3867$ & $1.8323$ & $2.4262$ & 1.98009(81) & 5.362(20) & 1.003(16) & 0.01799(73) & 48.7(2.0) \\
  \hline
\end{tabular}

\vspace{3ex}

 \begin{tabular}{|lllllllll|}
 \hline
    \rowcolor{gray!15} Ens.  & $\wm am_Q$ & $\nu$ & $c_{E,B}$  & $a E_{D_s}$       & $E_{D_s}$ [GeV]   & $c^2_{D_s}$       &  $a E_{D^*_s}-a E_{D_s}$ &  $E_{D^*_s}-E_{D_s}$ [MeV]\!\!\!\null  \\
  \hline
   CP & $\wm0.14752$        &  1.1928  &  1.7784 & 1.12144(43) & 1.9395(43)   & 1.0075(52)   &  0.07903(59)  &  136.7(1.1)  \\
   CP & $\wm0.27077$        &  1.1879  &  2.0714 & 1.13419(43) & 1.9616(44)   & 1.0008(51)   &  0.08353(63)  &  144.5(1.1)  \\
     \rowcolor{green!20}   CP & $\wm0.27514$        & $1.1883$ & $2.0712$ & 1.13804(43) & 1.9682(44)   & 1.0000(51)  &  0.08317(63)  &  143.8(1.1)  \\
  \hline
   C005   & $\wm0.11360$    &  1.2045  &  1.7936 & 1.08504(53) &  1.9366(55)  & 1.0078(95)     &  0.08152(80)  &  145.5(1.5)  \\ 
   C005   & $\wm0.11360$    &  1.2045  &  1.72   & 1.10561(55) &  1.9733(56)  & 1.0019(97)   &  0.07796(79)  &  139.1(1.5)  \\
   C005   & $\wm0.11360$    &  1.13    &  1.7936 & 1.04096(53) &  1.8579(53)  & 0.9646(89)   &  0.08375(82)  &  149.5(1.5)  \\
   C005   & $\wm0.14$       &  1.2045  &  1.7936 & 1.10676(55) &  1.9753(56)  & 1.0018(96)  &  0.07934(80)  &  141.6(1.5)  \\
     \rowcolor{green!20}   C005   & $\wm0.15410$    & $1.2004$ & $1.8407$ & 1.10274(54) &  1.9682(56)  & 1.0001(95)   &  0.08057(81)  &  143.8(1.5)  \\
  \hline
   F004   & $-0.01750$      &  1.1039  &  1.5884 & 0.81570(39) & 1.9441(71)   & 1.003(10)   &  0.06339(84)  &  151.1(2.1)  \\
   F004   & $-0.01750$      &  1.1039  &  1.51   & 0.84109(41) & 2.0046(73)   & 0.999(11)     &  0.05992(83)  &  142.8(2.0)  \\
   F004   & $-0.01750$      &  1.05    &  1.5884 & 0.78075(39) & 1.8608(68)   & 0.9674(99)   &  0.06513(85)  &  155.2(2.1)  \\
   F004   & $\wm0.01$       &  1.1039  &  1.5884 & 0.84309(40) & 2.0093(73)   & 0.998(11)    &  0.06090(84)  &  145.1(2.1)  \\
     \rowcolor{green!20}   F004   & $-0.05167$      & $1.1021$ & $1.4483$ & 0.82583(40) & 1.9682(72)   & 1.000(11)    &  0.06032(82)  &  143.8(2.0)  \\
  \hline
   F1M    & $-0.01508$  & 1.0635   & 1.6094    & 0.72535(35) & 1.9643(73) & 0.9766(91) & 0.05356(84) & 145.0(2.3) \\ 
   F1M    & $\wm0.005$  & 1.0635   & 1.6094    & 0.74696(38) & 2.0228(75) & 0.972(10)  & 0.05181(86) & 140.3(2.4) \\ 
   F1M    & $-0.01508$  & 1.09     & 1.6094    & 0.74250(37) & 2.0107(75) & 0.994(10)  & 0.05274(84) & 142.8(2.3) \\ 
   F1M    & $-0.01508$  & 1.0635   & 1.7       & 0.69066(29) & 1.8703(69) & 0.9844(80) & 0.05757(82) & 155.9(2.3) \\ 
     \rowcolor{green!20}   F1M    & $-0.05874$  & $1.0941$ & $1.5345$  & 0.72459(34) & 1.9622(73) & 1.0007(93) & 0.05339(81) & 144.6(2.3) \\
  \hline
\end{tabular}
\end{center}

\vspace{-3ex}

\caption{\label{tab:RHQparams}Tuning of the bottom-quark action (upper table) and charm-quark action (lower table). Shown here are results obtained from fits to $B_s^{(*)}$ and $D_s^{(*)}$ two-point correlation functions for several choices of the action parameters, on four of the ensembles used. All energies listed here are at zero momentum. The final values of the action parameters, which satisfy the tuning conditions and were used for the new $\Lambda_b \to p, \Lambda, \Lambda_c$ form-factor calculations, are highlighted in green. The parameter choices highlighted in blue correspond to the central values of the bottom-quark parameters used by RBC/UKQCD \cite{Flynn:2023nhi}. The uncertainties on the bottom-quark parameters given in Ref.~\cite{Flynn:2023nhi} are not propagated here.}

\end{table}

\begin{figure}

\begin{center}

 \includegraphics[width=0.8\linewidth]{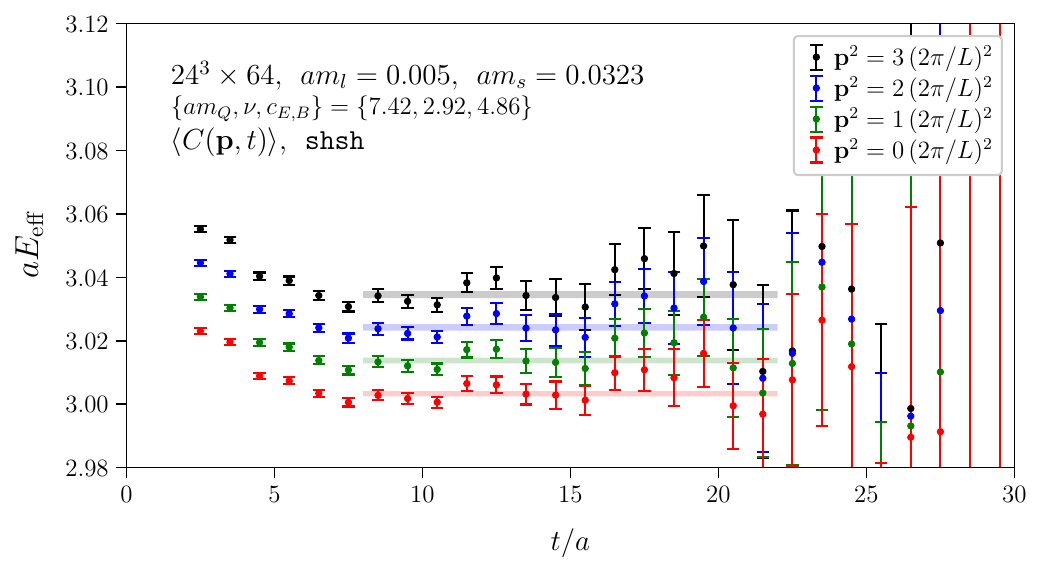}

\end{center}

\vspace{-3ex}

\caption{\label{fig:BsEeff}Effective energies of $B_s$ two-point functions used for the tuning of the bottom-quark parameters.}
 
\end{figure}

Note that the RBC and UKQCD collaborations also updated their tuning \cite{RBC:2012pds} of the bottom-quark parameters using the new lattice-spacing determinations from Ref.~\cite{RBC:2014ntl} and shared their results with me; these new parameters have since been published in Ref.~\cite{Flynn:2023nhi}. I tested the central values of the parameters determined by RBC/UKQCD, as shown in the columns highlighted in blue in Table \ref{tab:RHQparams}. I found that, while the $B_s$ energy at zero momentum and the hyperfine splitting agree with experiment, the $c^2$ values are 8\% and 14\% below 1 on the C005 and F004 ensembles, respectively (Ref.~\cite{Flynn:2023nhi} also presents uncertainties on the heavy-quark parameters which I did not propagate here). In contrast, RBC/UKQCD obtained $c^2$ equal to 1. My proposed explanation for this discrepancy is the following. Due to the particular choice of smearing scheme in Ref.~\cite{RBC:2012pds}, the excited-state contamination in the two-point functions shows a strong momentum dependence, as can be seen in Fig.~4 of Ref.~\cite{RBC:2012pds}. When computing the very small differences in $(aE)^2$ between zero and nonzero momentum, there is a large cancellation in statistical uncertainties due to correlations. However, the excited-state contamination in $(aE)^2$ changes sign from negative to positive as $\mathbf{p}^2$ is increased. Thus, the excited-state contamination in the energy difference or slope is {\it enhanced} and leads to an overestimate of $c^2$. I used a different smearing scheme for which the excited-state contamination in the two-point functions is almost momentum-independent, as shown in Fig.~\ref{fig:BsEeff}.

\begin{figure}

\begin{center}
 
 \includegraphics[width=0.8\linewidth]{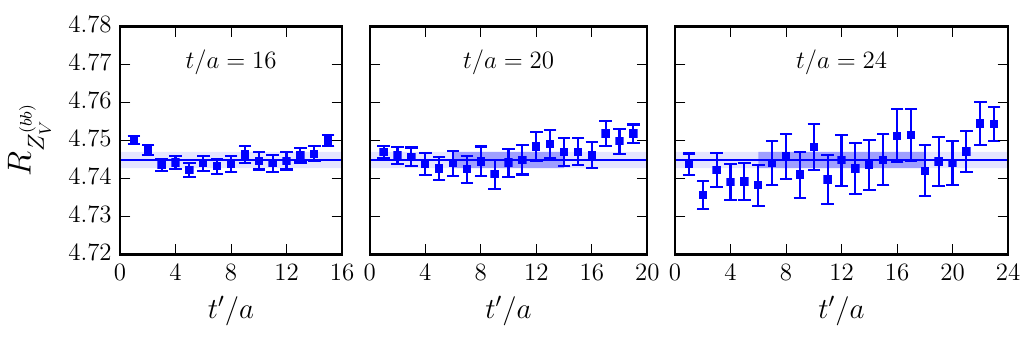}

\end{center}

\vspace{-2ex}

\caption{\label{fig:ZVbb}Numerical results for the ratio of zero-momentum $B_s$ two-point and three-point functions (with the current $\bar{b}\gamma^0 b$) that gives $Z_V^{bb}$, from the F004 ensemble. Here, $t$ is the source-sink separation and $t^\prime$ is the current insertion time. The dark-shaded regions are included in the constant fit.}
 
\end{figure}

The 2015/2016 calculations of the $\Lambda_b$-decay form factors used the ``mostly nonperturbative'' method \cite{Hashimoto:1999yp,ElKhadra:2001rv} to renormalize and $\mathcal{O}(a)$-improve the weak currents. The renormalized and improved $b\to q$ currents ($q=u,s,c$) are of the form
\begin{equation}
 J_\Gamma = \sqrt{ Z_V^{qq} \:Z_V^{bb}}  \:\:\rho_\Gamma\:\Big[ \bar{q} \, \Gamma\, b \:+\: \mathcal{O}(a)\text{ improvement terms} \Big],
\end{equation}
where $Z_V^{bb}$, $Z_V^{qq}$ are the renormalization factors of $\bar{q}\gamma^0 q$ and  $\bar{b}\gamma^0 b$, computed nonperturbatively using charge conservation, and the residual matching coefficients $\rho_\Gamma$ as well as the $\mathcal{O}(a)$ improvement coefficients are computed to one loop in lattice perturbation theory \cite{Detmold:2015aaa} (tree-level only for the tensor currents). For the preliminary analysis of the next-generation form-factor data shown in the following, I use the same method. I newly computed $Z_V^{bb}$ and $Z_V^{cc}$ for the updated heavy-quark parameters from Table \ref{tab:RHQparams} using ratios of zero-momentum $B_s$/$D_s$ two-point and three-point functions. Example plots of these ratios are shown in Fig.~\ref{fig:ZVbb}. My results for $Z_V^{cc}$ and $Z_V^{bb}$, along with the values I currently use for $Z_V^{ll}$ ($l=u,s$), are given in Table \ref{tab:ZV}. For the residual-matching and $O(a)$-improvement coefficients in the preliminary analysis, I still use the values from Ref.~\cite{Detmold:2015aaa}, which were computed for the old heavy-quark-action parameters. For the CP ensemble, for now, I use the same values as for C005. For the F1M ensemble, I extrapolated the values from Ref.~\cite{Detmold:2015aaa} linearly in the lattice spacing.

\begin{table}
\begin{center}\footnotesize
\begin{tabular}{|llll|}
\hline
   \rowcolor{gray!10} Ensemble          &  $Z_V^{ll}$  & $Z_V^{cc}$    &  $Z_V^{bb}$     \\
\hline
  CP            &  $0.71076(25)$ \cite{RBC:2014ntl}                  &  $1.40756(17)$   & $9.9128(81)$       \\[0.2ex]
  C005, C005LV  &  $0.71273(26)$ \cite{RBC:2014ntl}    &  $1.35761(16)$   & $9.0631(84)$  \\[0.2ex]
  F004, F006    &  $0.7440(18)$  \cite{RBC:2014ntl}      &  $1.160978(74)$  & $4.7449(21)$  \\[0.2ex]
  F1M           &  $0.761125(19)$ ($Z_A$, F1S) \cite{Boyle:2017jwu}    &  $1.112316(61)$  & $3.7777(23)$  \\[0.2ex]
\hline
\end{tabular}
\end{center}
\caption{\label{tab:ZV} My results for the nonperturbative renormalization factors $Z_V^{cc}$ and $Z_V^{bb}$, along with the values I currently use for $Z_V^{ll}$ ($l=u,s$). For the F1M ensemble, $Z_V^{ll}$ was not yet available, so in this preliminary analysis I used the value $Z_A^{ll}$ from the F1S ensemble \cite{Boyle:2017jwu} that differs from F1M only by having the Shamir domain-wall action instead of the M\"obius action \cite{Boyle:2018knm}.}
\end{table}

\begin{figure}

\hspace{6ex} \includegraphics[height=0.0215\textheight]{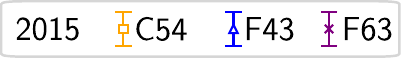} 
\hspace{4.2ex} \includegraphics[height=0.021\textheight]{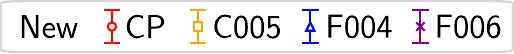} 
\hspace{6.2ex} \includegraphics[height=0.0215\textheight]{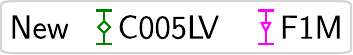}

\includegraphics[width=0.33\linewidth]{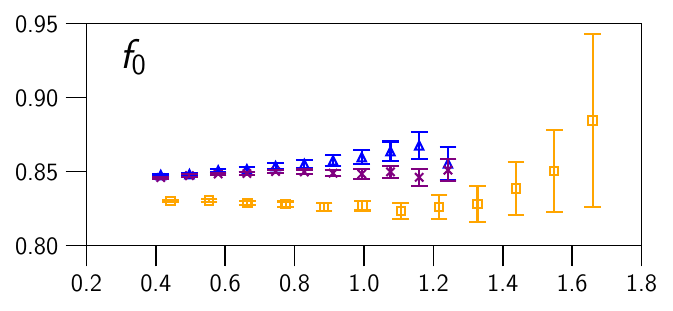}  \includegraphics[width=0.33\linewidth]{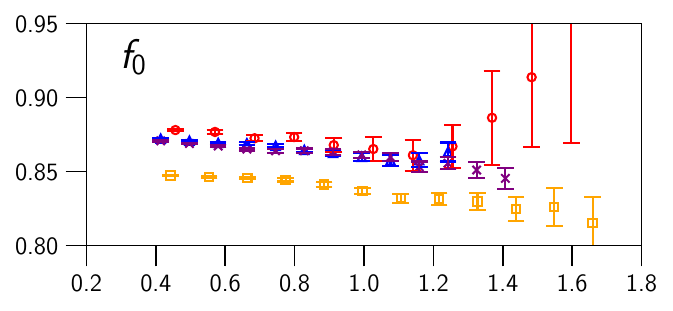} \includegraphics[width=0.33\linewidth]{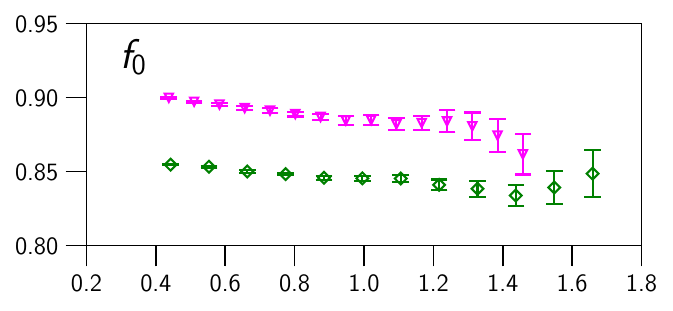}

\includegraphics[width=0.33\linewidth]{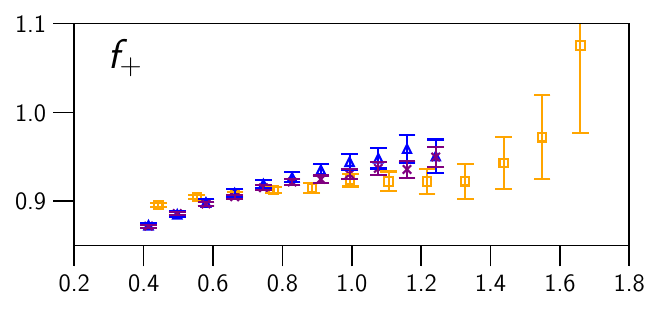}  \includegraphics[width=0.33\linewidth]{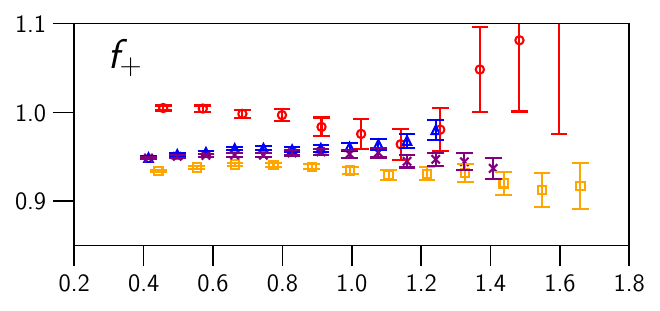} \includegraphics[width=0.33\linewidth]{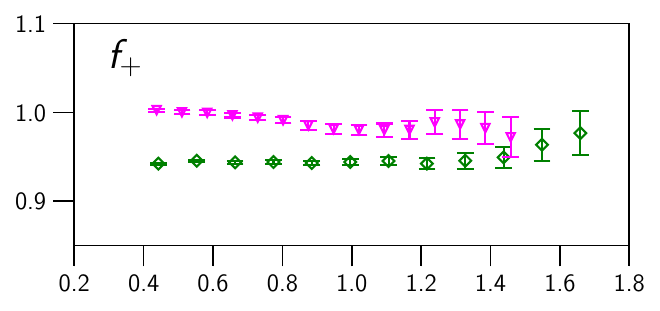}

\includegraphics[width=0.33\linewidth]{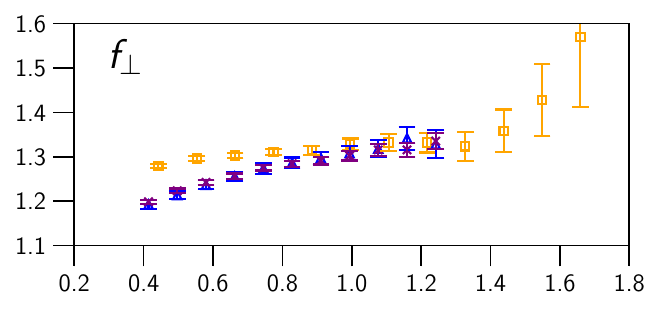}  \includegraphics[width=0.33\linewidth]{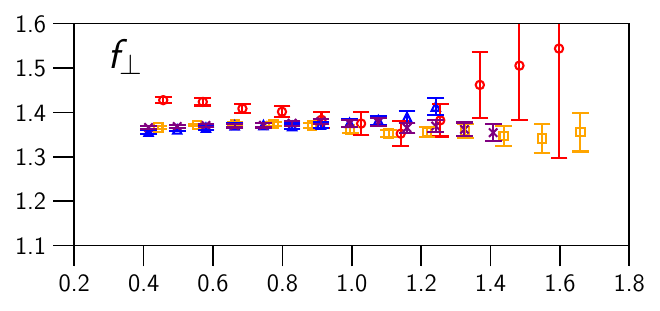} \includegraphics[width=0.33\linewidth]{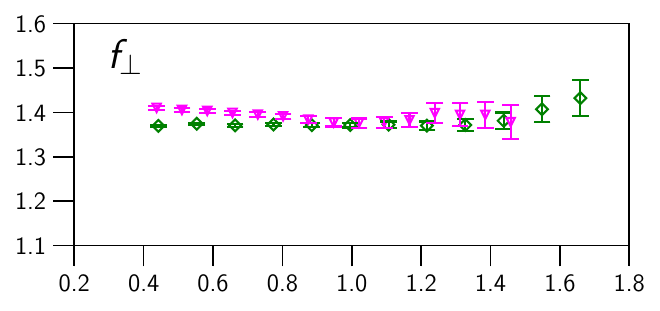}

\includegraphics[width=0.33\linewidth]{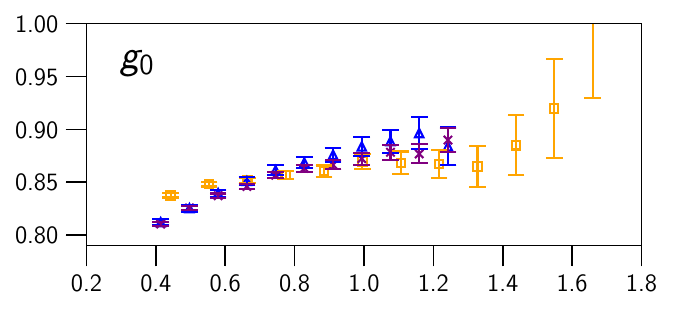}  \includegraphics[width=0.33\linewidth]{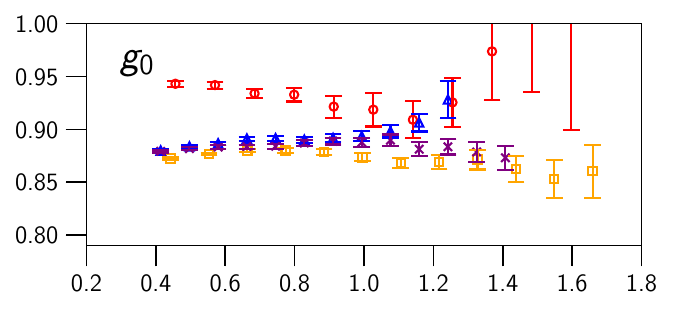} \includegraphics[width=0.33\linewidth]{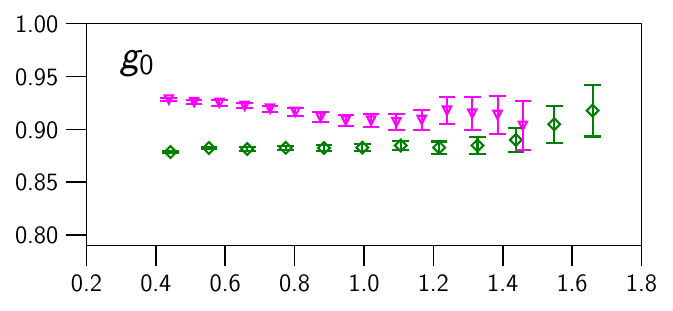}

\includegraphics[width=0.33\linewidth]{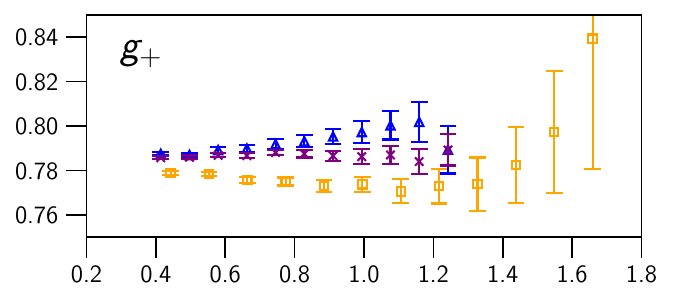}  \includegraphics[width=0.33\linewidth]{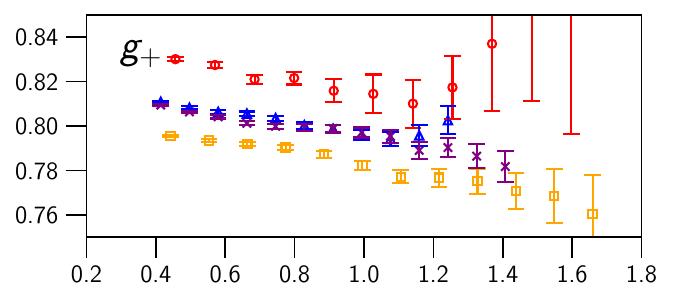} \includegraphics[width=0.33\linewidth]{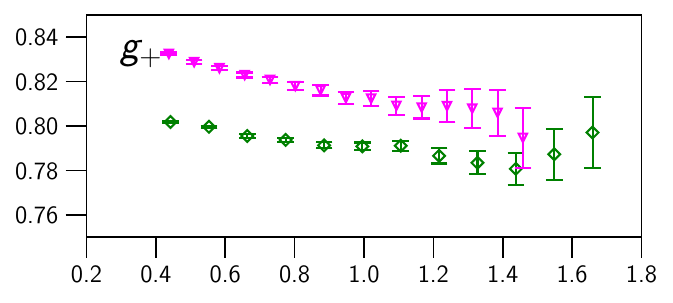}

\includegraphics[width=0.33\linewidth]{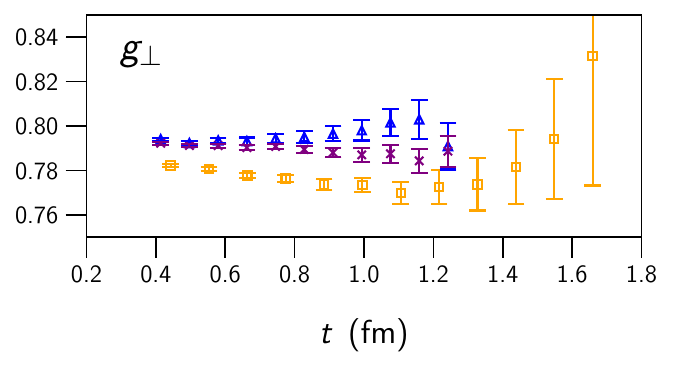}  \includegraphics[width=0.33\linewidth]{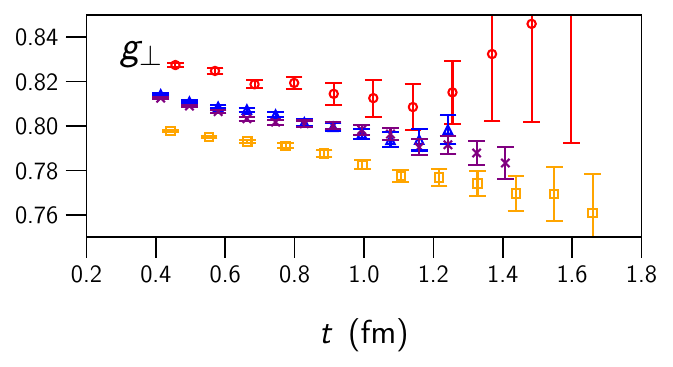} \includegraphics[width=0.33\linewidth]{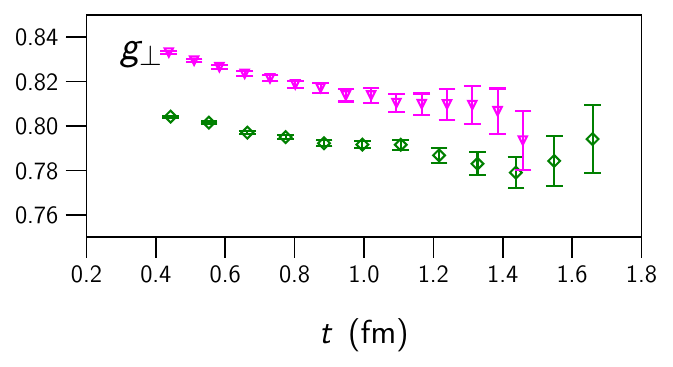}

\caption{\label{fig:LbLc}The quantities $R_f(|\mathbf{p}^\prime|,t)$ that, for large $t$, become equal to the $\Lambda_b \to \Lambda_c$ vector and axial-vector form factors at $|\mathbf{p}^\prime|\approx 0.8\:{\rm GeV}$.  The center and right plots are from the next-generation computations (with preliminary renormalization factors); the left plots show the corresponding 2015 results for comparison. }
\end{figure}

\begin{figure}
\hspace{3ex} \includegraphics[height=0.0215\textheight]{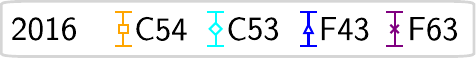} 
\hspace{3.2ex} \includegraphics[height=0.021\textheight]{figures/formfactors/legend_CP_C005_F004_F006.pdf} 
\hspace{6.2ex} \includegraphics[height=0.0215\textheight]{figures/formfactors/legend_C005LV_F1M.pdf}

\includegraphics[width=0.33\linewidth]{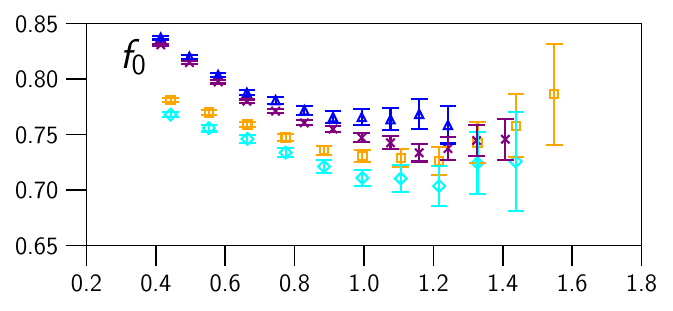}  \includegraphics[width=0.33\linewidth]{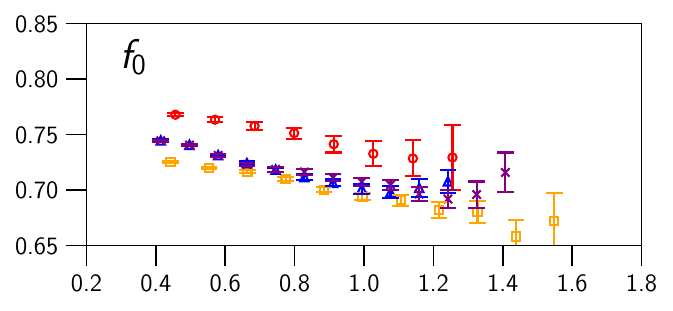} \includegraphics[width=0.33\linewidth]{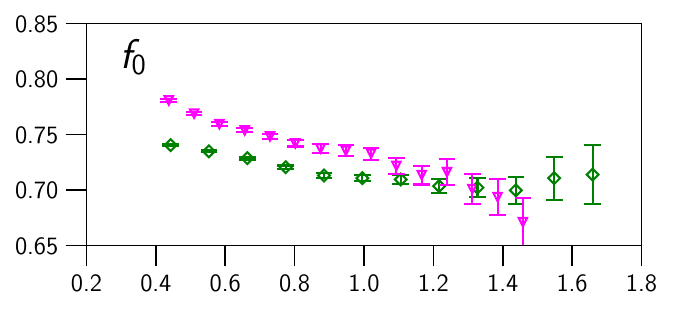}

\includegraphics[width=0.33\linewidth]{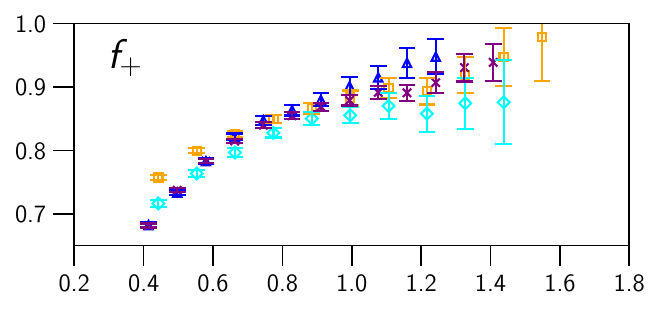}  \includegraphics[width=0.33\linewidth]{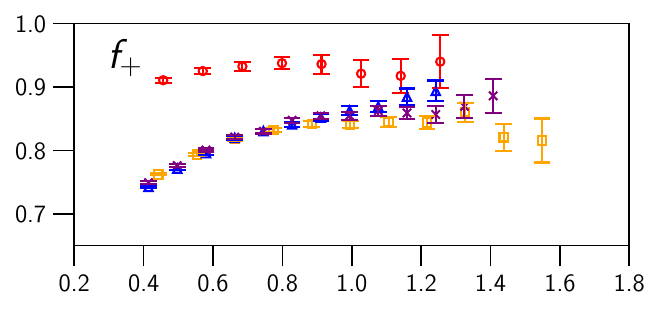} \includegraphics[width=0.33\linewidth]{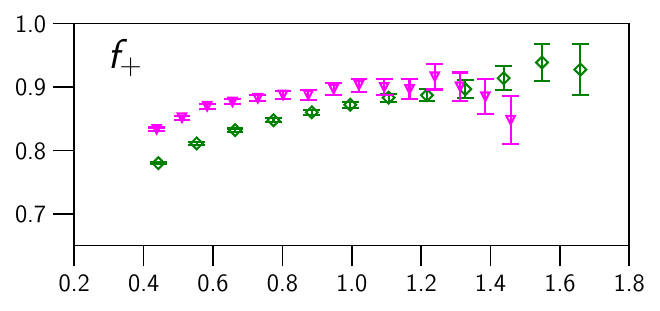}

\includegraphics[width=0.33\linewidth]{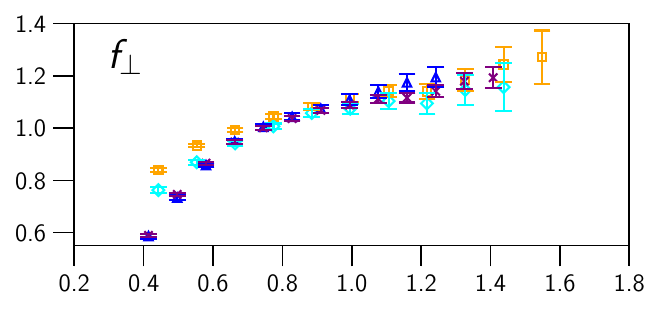}  \includegraphics[width=0.33\linewidth]{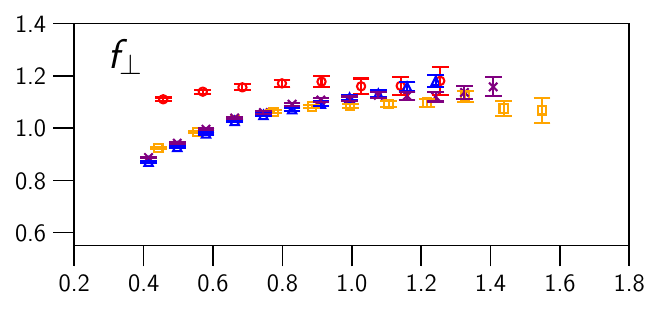} \includegraphics[width=0.33\linewidth]{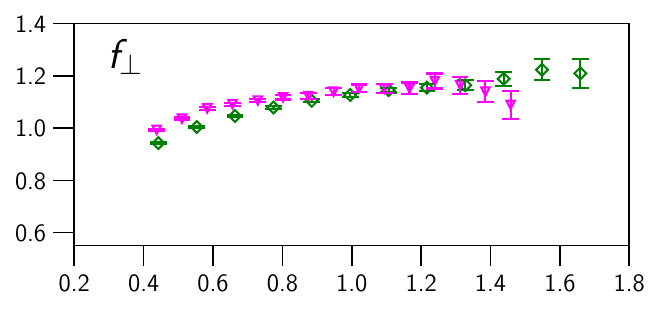}

\includegraphics[width=0.33\linewidth]{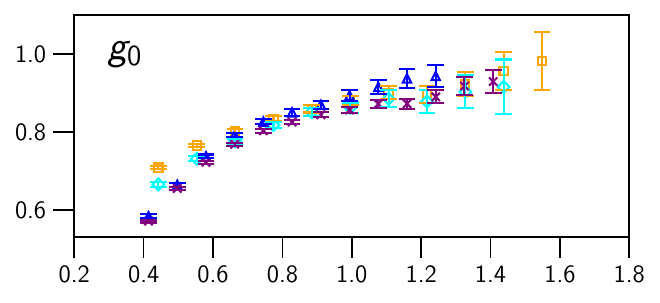}  \includegraphics[width=0.33\linewidth]{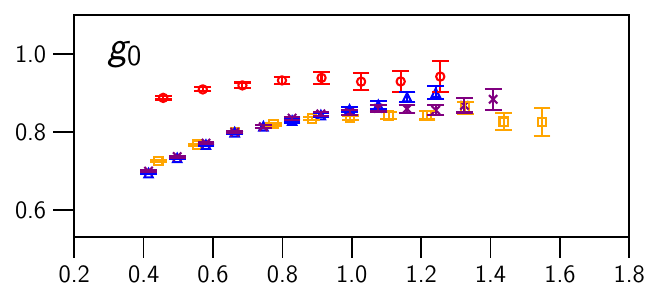} \includegraphics[width=0.33\linewidth]{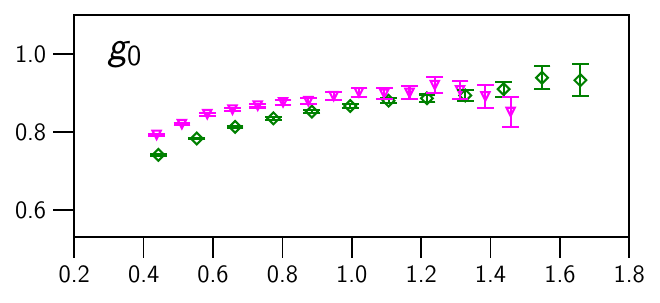}

\includegraphics[width=0.33\linewidth]{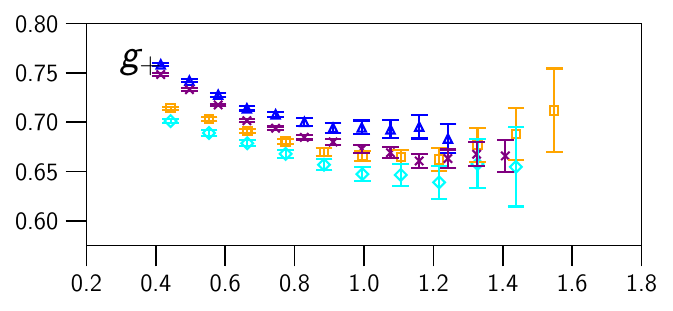}  \includegraphics[width=0.33\linewidth]{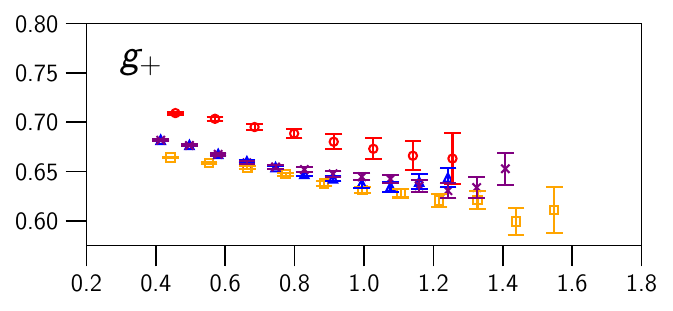} \includegraphics[width=0.33\linewidth]{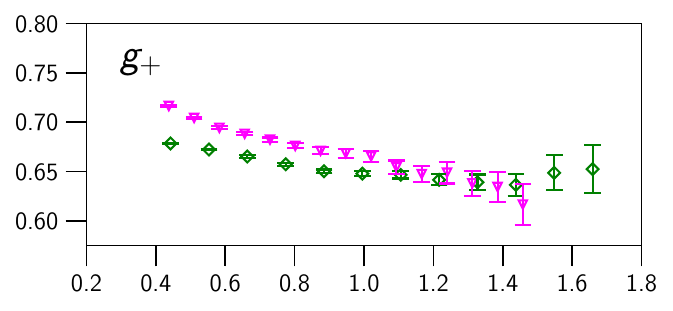}

\includegraphics[width=0.33\linewidth]{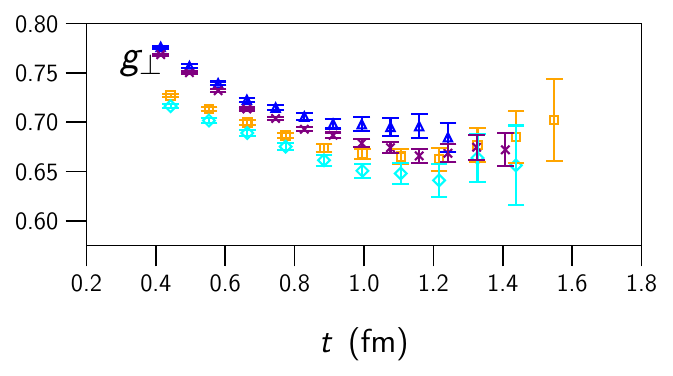}  \includegraphics[width=0.33\linewidth]{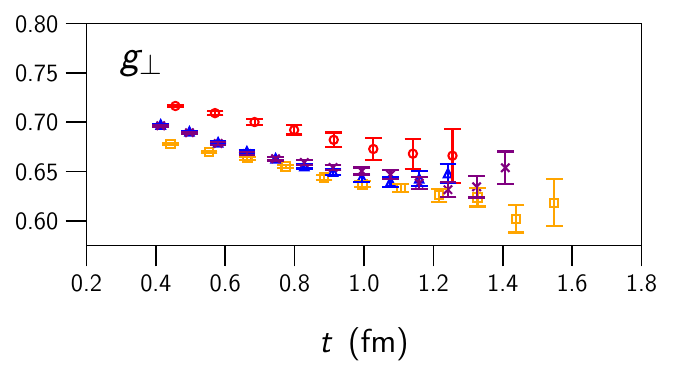} \includegraphics[width=0.33\linewidth]{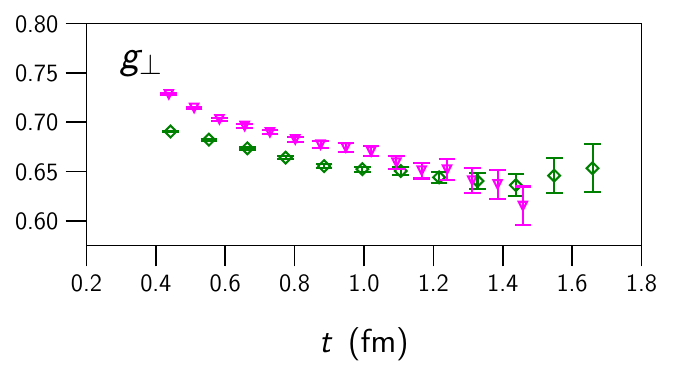}

\caption{\label{fig:LbL}Like Fig.~\protect\ref{fig:LbLc}, but for $\Lambda_b \to \Lambda$. In comparing the old and new results, note that the valence strange-quark masses have also changed (see Table \protect\ref{tab:ensembles}). }

\end{figure}

\begin{figure}
\hspace{6ex} \includegraphics[height=0.0215\textheight]{figures/formfactors/legend_C54_F43_F63.pdf} 
\hspace{4.2ex} \includegraphics[height=0.021\textheight]{figures/formfactors/legend_CP_C005_F004_F006.pdf} 
\hspace{6.2ex} \includegraphics[height=0.0215\textheight]{figures/formfactors/legend_C005LV_F1M.pdf}

\includegraphics[width=0.33\linewidth]{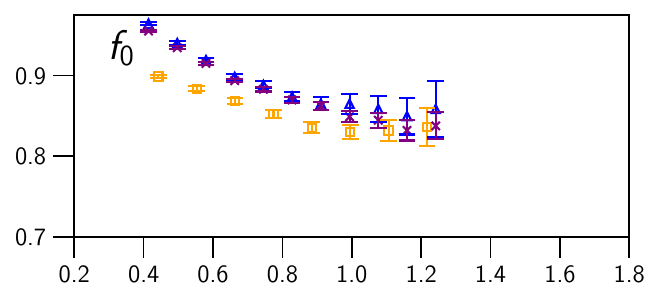}  \includegraphics[width=0.33\linewidth]{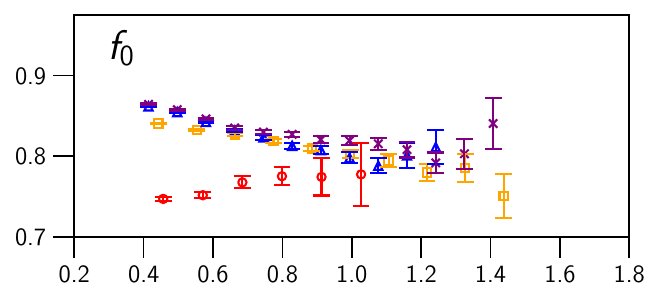} \includegraphics[width=0.33\linewidth]{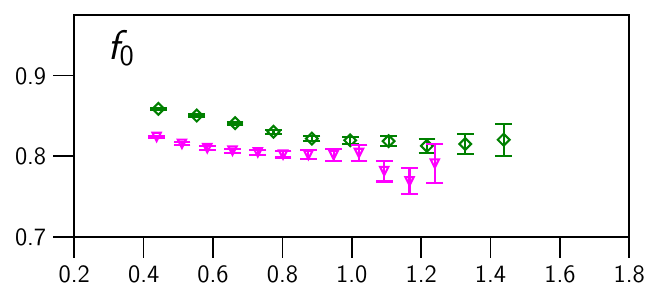}

\includegraphics[width=0.33\linewidth]{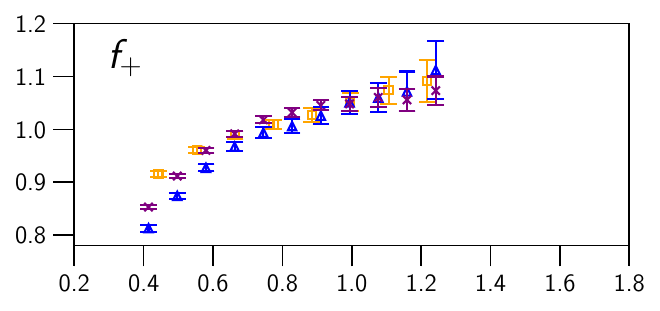}  \includegraphics[width=0.33\linewidth]{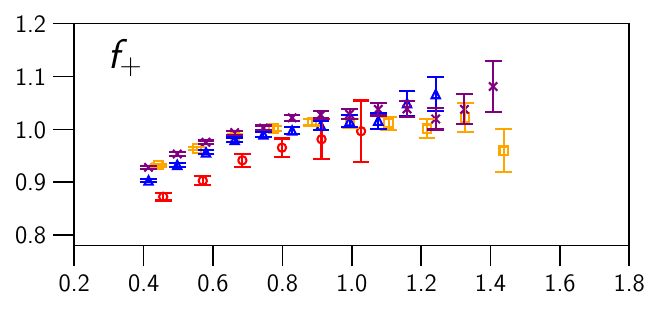} \includegraphics[width=0.33\linewidth]{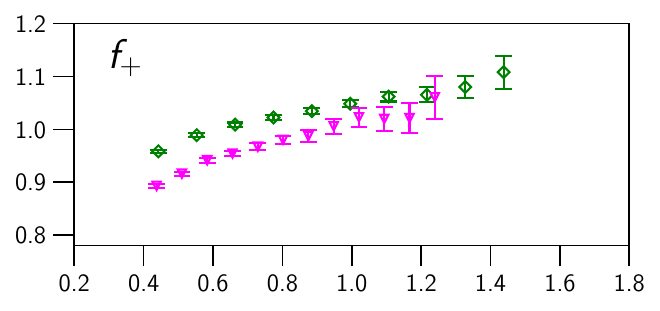}

\includegraphics[width=0.33\linewidth]{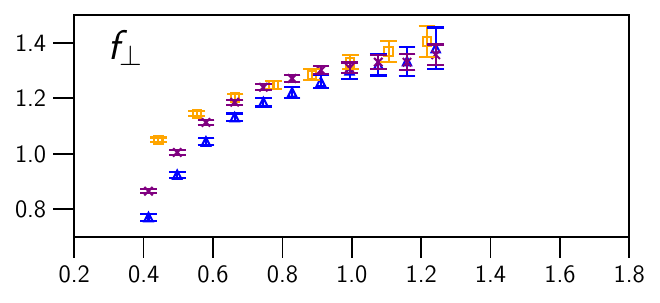}  \includegraphics[width=0.33\linewidth]{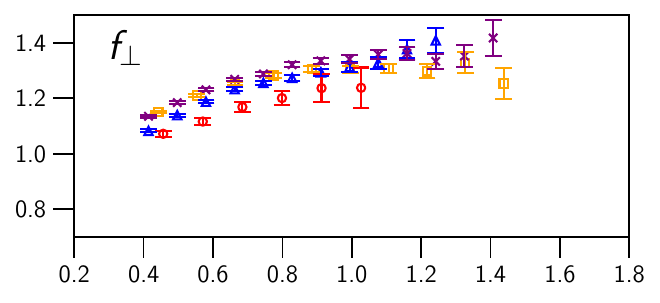} \includegraphics[width=0.33\linewidth]{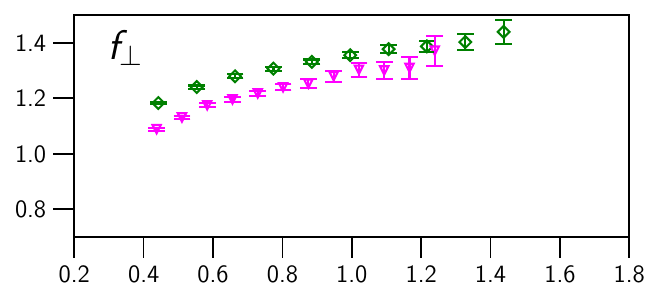}

\includegraphics[width=0.33\linewidth]{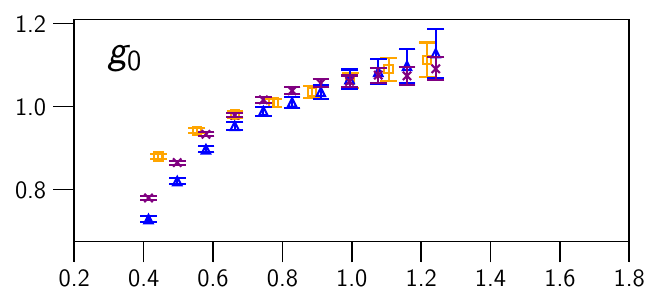}  \includegraphics[width=0.33\linewidth]{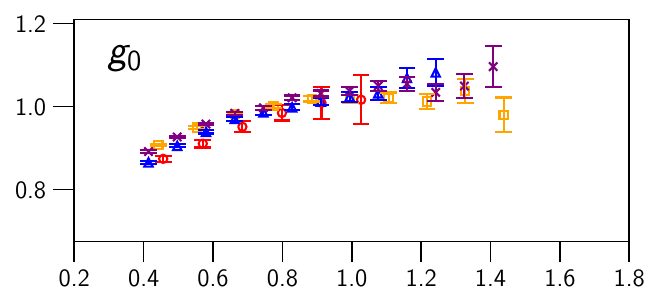} \includegraphics[width=0.33\linewidth]{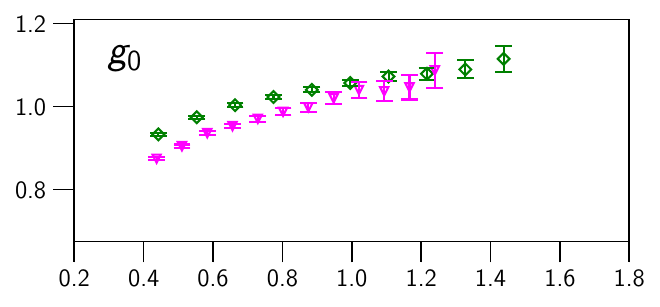}

\includegraphics[width=0.33\linewidth]{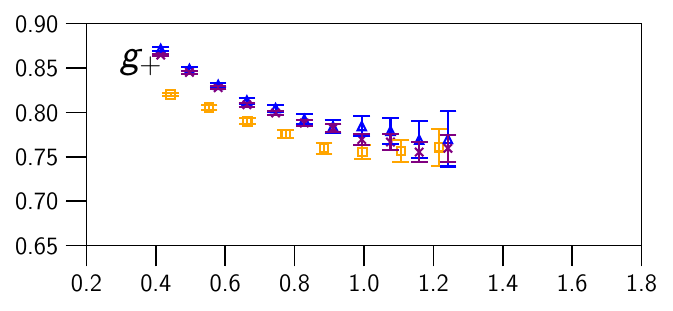}  \includegraphics[width=0.33\linewidth]{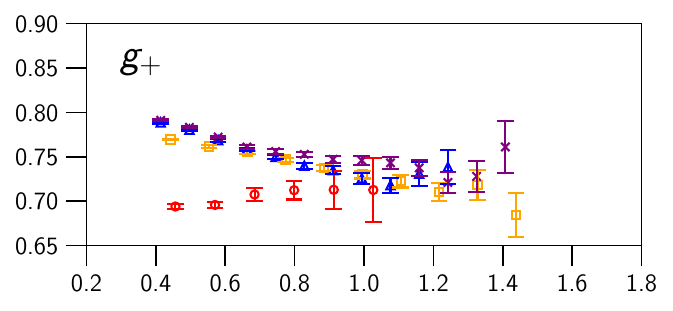} \includegraphics[width=0.33\linewidth]{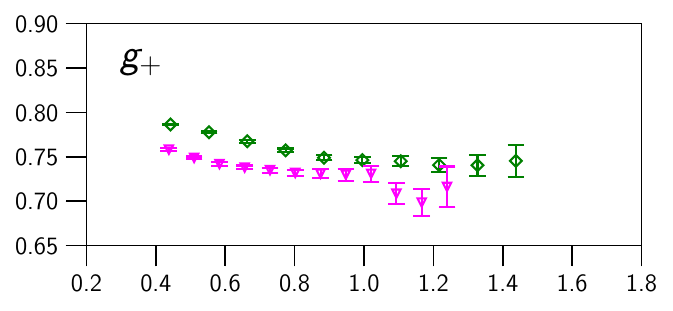}

\includegraphics[width=0.33\linewidth]{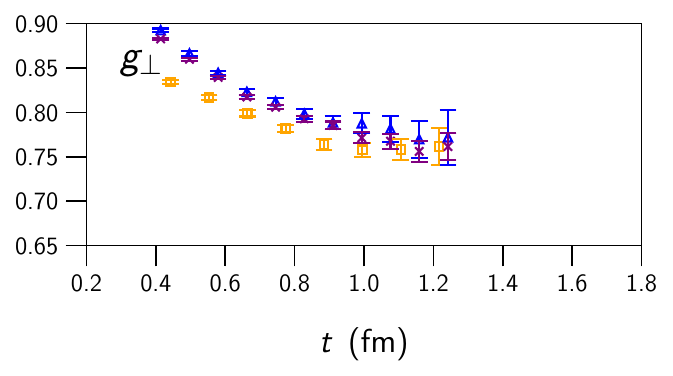}  \includegraphics[width=0.33\linewidth]{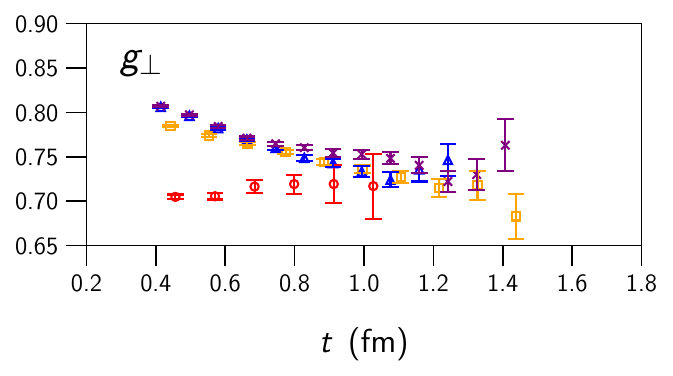} \includegraphics[width=0.33\linewidth]{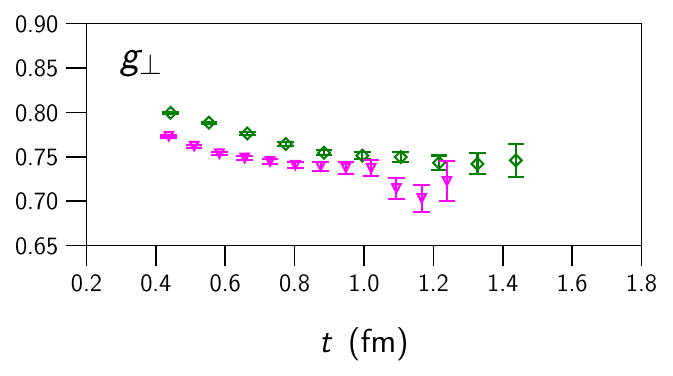}

 \caption{\label{fig:Lbp}Like Fig.~\protect\ref{fig:LbLc}, but for $\Lambda_b \to p$. }
 
\end{figure}

\section{Preliminary form-factor results}

The form factors are obtained from products of forward and backward $\Lambda_b\to F$ ($F=p,\Lambda,\Lambda_c$) three-point functions divided by a product of the $\Lambda_b$ and $F$ two-point functions; in these ratios, the overlap factors and time dependence cancel for the ground-state contribution \cite{Detmold:2015aaa}. The calculations are done in the $\Lambda_b$ rest frame, for final-state momenta in the range $0<|\mathbf{p}^\prime|\lesssim 1.5$ GeV, and for all possible source-sink separations in the range $0.4\:{\rm fm}\lesssim t \lesssim 1.7\:{\rm fm}.$ The ratio is evaluated at current-insertion time $t^\prime=t/2$, where excited-state contamination is expected to be minimal. By multiplying with the appropriate kinematic factors and taking the square root, quantities $R_f(|\mathbf{p}^\prime|,t)$ are obtained that are equal to the form factor $f(|\mathbf{p}^\prime|)$ up to excited-state contamination that decays exponentially with $t$ \cite{Detmold:2015aaa}. I have performed preliminary fits of the $t$ dependence to extract the ground-state form factors, but I am still working on the estimates of the uncertainties. For this reason, I only present the results for $R_f(|\mathbf{p}^\prime|,t)$ here, at one kinematic point given by
\begin{equation}
  |\mathbf{p^\prime}|^2 = \left\{ \begin{array}{ll} 3 \cdot (2\pi/L)^2 & {\rm for\:\: C005, F004, F006}, \\ 12 \cdot (2\pi/L)^2 & {\rm for\:\: CP}, \\ 5 \cdot (2\pi/L)^2 & {\rm for\:\: C005LV, F1M} \\ \end{array}\right \} \approx (0.8\:{\rm GeV})^2.
\end{equation}
These results are shown in Fig.~\ref{fig:LbLc} for $\Lambda_b \to \Lambda_c$, Fig.~\ref{fig:LbL} for $\Lambda_b \to \Lambda$, and Fig.~\ref{fig:Lbp} for $\Lambda_b \to p$. The corresponding results from the 2015/2016 data sets with $m_\pi^{(\rm val)}=m_\pi^{(\rm sea)}$ are included for comparison. Note that, at the larger source-sink separations, the fluctuations in the ratio may become too large to compute the square root; such separations are omitted.

The results for $\Lambda_b \to \Lambda_c$ have the smallest statistical uncertainties. For some form factors, even after the large-$t$ extrapolation, the dependence on the pion mass and lattice spacing will be clearly resolved with the help of the new CP and F1M data. For $\Lambda_b \to p$, it appears that the dependence on pion mass and lattice spacing will typically not be significant after extracting the ground-state contributions, in part due to the larger statistical uncertainties. Another feature worth pointing out is that, except for $g_+(\Lambda_b\to\Lambda_c)$ and $g_\perp(\Lambda_b\to\Lambda_c)$, the new F004 and F006 data generally show less excited-state contamination (i.e., smaller $t$-dependence) than the old F43 and F63 data due to the increased smearing width.

\section{Next steps}

The next task is to complete the fits of the $t$ dependence of $R_f(|\mathbf{p}^\prime|,t)$ to extract the ground-state form factors for all data sets and all momenta, after finalizing the values of the residual matching factors $\rho_\Gamma$ and the $\mathcal{O}(a)$-improvement coefficients in the weak currents. It may also be possible to replace the mostly nonperturbative renormalization method with a fully nonperturbative method \cite{Giusti:2021rsf}. Finally, the chiral/continuum/kinematic extrapolations of the form factors need to be performed. In the kinematic extrapolations, I plan to consider dispersive bounds, whose application to baryon semileptonic form factors using a novel parametrization was discussed in Ref.~\cite{Blake:2022vfl}. The parametrization replaces the usual $z^n$ monomials by polynomials $p_n(z)$ that are orthonormal on the arc of the unit circle that is relevant for the dispersive bounds \cite{Gubernari:2020eft}.

\section*{Acknowledgments}

I thank the RBC and UKQCD collaborations for providing the gauge-field ensembles. I thank Oliver Witzel for sharing preliminary values of the bottom-quark action parameters with me prior to their publication. I am currently supported by the U.S. Department of Energy, Office of Science, Office of High Energy Physics under Award Number D{E-S}{C0}009913, and was supported by National Science Foundation Award Number PHY-1520996 and by the RHIC Physics Fellow Program of the RIKEN BNL Research Center during the early stages of this work. I carried out the computations on facilities at the National Energy Research Scientific Computing Center, a DOE Office of Science User Facility supported by the Office of Science of the U.S. Department of Energy under Contract No.~DE-AC02-05CH1123, and on facilities of the Extreme Science and Engineering Discovery Environment (XSEDE), which was supported by National Science Foundation grant number ACI-1548562. I used Chroma \cite{Edwards:2004sx,Chroma}, QLUA \cite{QLUA}, MDWF \cite{MDWF}, and related USQCD software \cite{USQCD}.

\FloatBarrier

\providecommand{\href}[2]{#2}\begingroup\raggedright\endgroup

\end{document}